\documentclass[final,5p,times,twocolumn]{elsarticle}
\usepackage{graphicx}  
\usepackage{dcolumn}   
\usepackage{bm}        
\usepackage{amssymb}   
\usepackage{graphicx, epsfig, subfigure}
\graphicspath{{tmpfig/}{figures/}}

\usepackage{lipsum,amsmath,multicol}
\usepackage{subfigure}
\usepackage{mathrsfs}
\usepackage{longtable}
\usepackage{rotating}
\usepackage{mathtools}
\usepackage{cuted}
\usepackage{color}
\usepackage{lineno}
\usepackage{url}

\usepackage{setspace}
\journal{NIMA}

\begin{document}
\begin{frontmatter}
\date{\today}

    \title{Compact Scintillator Array Detector (ComSAD) for sounding rocket and CubeSat missions } 
                
\author[ncku_phy]{Pu Kai Wang \corref{cor1}}
\author[ncku_phy]{Chih-Yun Chen} \corref{cor1}
\author[ncku_phy]{Hsiang-Chieh Hsu}
\author[ncku_phy]{Mu-Hsin Chang}
\author[ncku_phy]{Wei Tai Liu}
\author[ncku_isps]{Hui-Kuan Fang \corref{cor1}}
\author[ncku_phy]{Ting-Chou Wu}
\author[ncku_isps]{Wen-Hao Chen \corref{cor1}}
\author[ncku_ee]{Chin Cheng Tsai}
\author[ncku_phy]{Alfred Bing-Chih Chen}
\author[ncku_phy,ncku_ms]{Yi Yang\corref{cor2}} 
\ead{yiyang@ncku.edu.tw}

\address[ncku_phy]{Department of Physics, National Cheng Kung University, Tainan, 70101, Taiwan, ROC}
\address[ncku_isps]{Institute of Space and Plasma Sciences, National Cheng Kung University, Tainan, 70101, Taiwan, ROC}
\address[ncku_ee]{Department of Electrical Engineering, National Cheng Kung University, Tainan, 70101, Taiwan, ROC}
\address[ncku_ms]{Department of Mechanical Engineering, National Cheng Kung University, Tainan, 70101, Taiwan, ROC}
    
    \cortext[cor1]{P.K. Wang is now at Universite Paris-Saclay (France), C.Y. Chen is now at ASUSTek Computer Inc., H.K. Feng is now at Taiwan Semiconductor Manufacturing Company, and W.H.Chen is now at Taiwan Innovative Space.}
\cortext[cor2]{Corresponding author.}

\begin{abstract}
   The development of CubeSats and more frequent launch chances of sounding rockets are a total game changer to the space program, and it allows us to build space instruments that are technologically feasible and affordable. 
    Therefore, it gives us a good opportunity to build a small cosmic-ray detector which has capabilities to measure the flux, direction, and even energy of cosmic rays at an altitude above the limitation of balloon experiments, and it may open a new door for building a constellation of detectors to study cosmic-ray physics. 
    Compact Scintillator Array Detector (ComSAD) is a funded sounding rocket mission of Taiwan's National Space Organization. 
    In this paper, we present the concept, design, and performance of ComSAD which is also suitable for future CubeSat missions. 
\end{abstract}


\begin{keyword}
Cosmic ray, scintillator detector, CubeSat, sounding rocket.
\end{keyword}

\end{frontmatter}


\section{Introduction}
\begin{figure}[!htbp]
  \begin{center}
      \includegraphics[width=0.47\textwidth]{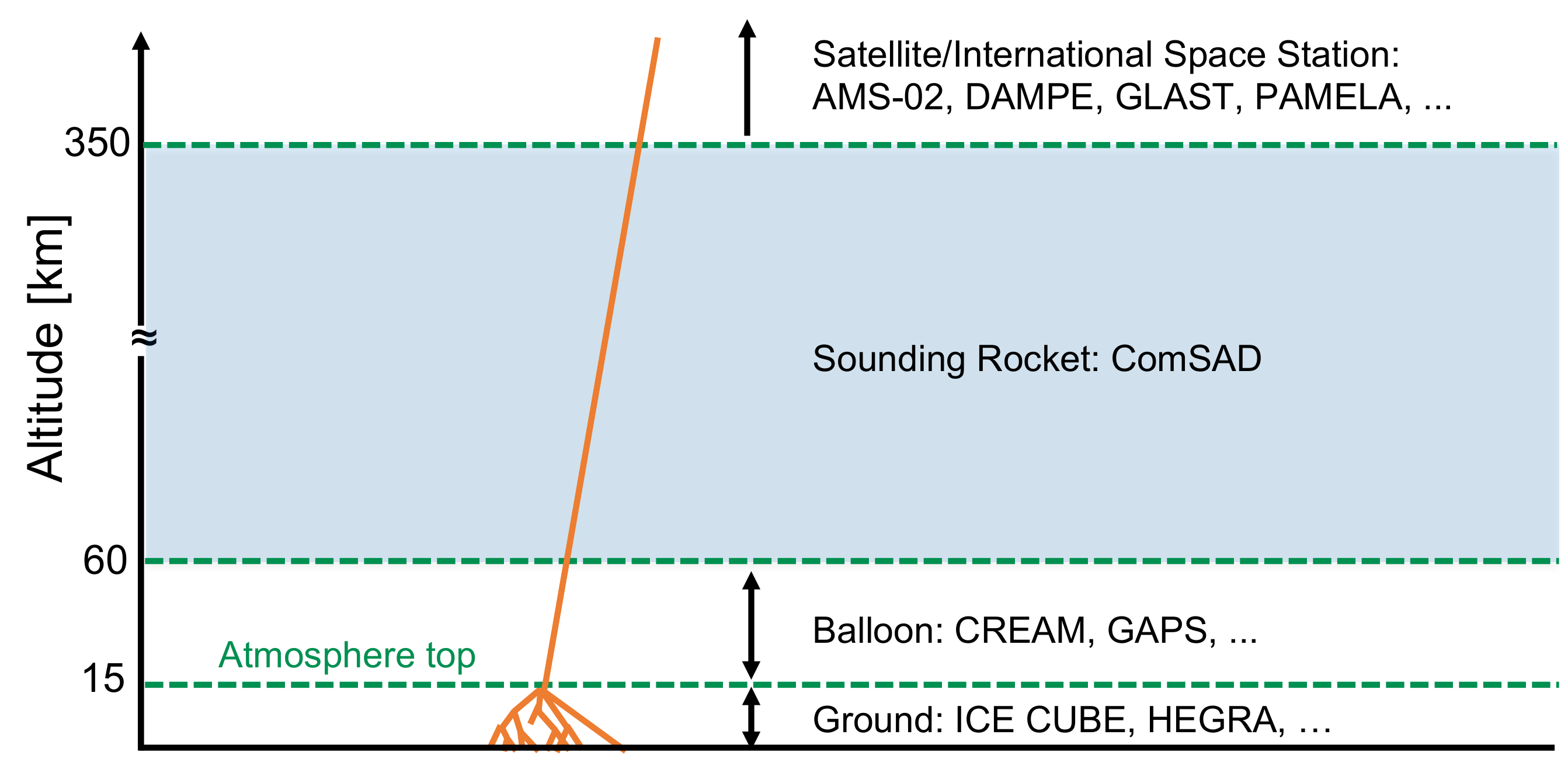}
      \end{center}
    \caption{Examples of cosmic ray experiments at different altitudes.
  \label{fig:cr_altitude}}
\end{figure}

Although cosmic rays have been discovered and studied for over one hundred years~\cite{cr_hist_1,cr_hist_2}, there are still many open questions to be answered, such as the nature of their origin and defining properties, for example. 
Cosmic rays are also a unique, natural source that allows one to study ultra-high energy physics. 
Since different physics can be probed at different altitudes, many experiments in the world are dedicated to cosmic-ray physics, and include huge array of detectors on or under ground, such as IceCube~\cite{icecube} and HEGRA~\cite{hegra}, and smaller detectors on balloons, such as CREAM~\cite{cream} and GAPS~\cite{gaps}, and in low-Earth orbit, like AMS-02~\cite{ams} and PAMELA~\cite{pamela}. 
However, there is a gap of the cosmic-ray measurements at the altitude between balloon and satellite experiments which can be filled by sounding rocket experiments~\cite{rocket}. 
Figure~\ref{fig:cr_altitude} shows some examples of different types of cosmic-ray experiments at different altitudes.   

Interestingly, there are no cosmic-ray measurements conducted simultaneously at different locations, except for the gravitational wave experiment~\cite{gw_exp}, and the recent development of the CubeSat missions~\cite{cubesat} that make simultaneous measurements of cosmic rays around Earth (a 4-dimensional map of cosmic rays) affordable, namely by building a constellation of detectors.   
There are many restrictions on the payload design due to the limitations on the weight, size, and data transmission rate for sounding rockets and CubeSats. 
Additionally, the failure rate of the sounding rocket and CubeSat missions is higher than that of normal space programs requiring that the cost be carefully controlled.
$\bf{Com}$pact $\bf{S}$cintillator $\bf{A}$rray $\bf{D}$etector (ComSAD) is designed specifically for the ``forward-looking hybrid sounding rocket project'' in Taiwan which is one of the major space programs in National Space Organization (NSPO)~\cite{nspo}.
With some minor modifications, ComSAD can also be suitable for the future CubeSat and aircraft missions. 

The paper is organized as follows: Section 2 describes the details of the ComSAD detector. Section 3 shows performance of ComSAD. Section 4 demonstrates modifications of ComSAD needed to make it portable and suitable for aircraft missions. Finally, the conclusions are given in Section 5.

\section{The ComSAD detector}
\label{sec:comsad}
\begin{figure}[!htbp]
  \begin{center}
      \includegraphics[width=0.48\textwidth]{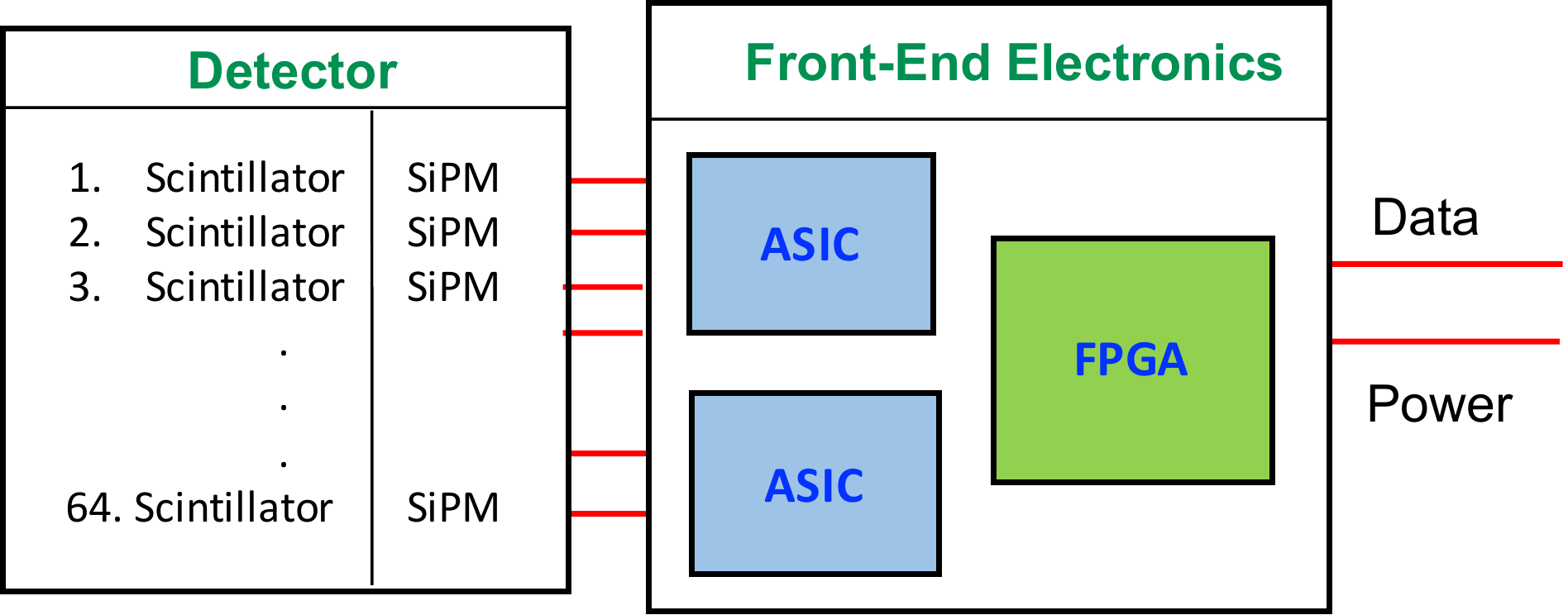}
      \end{center}
    \caption{The ComSAD system block diagram. 
  \label{fig:block_diagram}}
\end{figure}

The key components of the ComSAD detector are the scintillator, silicon photomultiplier (SiPM), application-specific integrated circuit (ASIC), field-programmable gate array (FPGA), power supply unit (PSU), and supporting structure.
The basic idea of ComSAD is to use the property of scintillators related to their intrinsic operation. 
Photons are emitted when a high-energy particle passes through the scintillator, and these photons can then be detected by a sensor.
Moreover, with a specific combination of scintillators, the direction of the incident particles can be determined. 
Due to the limitation of the weight on the sounding rocket mission, ComSAD is composed of a total of 64 scintillators, and each scintillator, which has the size of 1 $\times$ 1 $\times$ 4 $cm^3$, is paired with a SiPM to detect the photons. 
There are 2 ASICs controlled by an FPGA to handle the signals from the 64 SiPMs. 
The PSU converts the input 12 V DC, from the rocket or battery, to 5 V for the ASIC, and the DC-DC converter boosts the input to 29 V for the SiPM. 
Figure~\ref{fig:block_diagram} shows the block diagram for the ComSAD system. 

\begin{figure}[htbp]
  \begin{center}
      \includegraphics[width=0.45\textwidth]{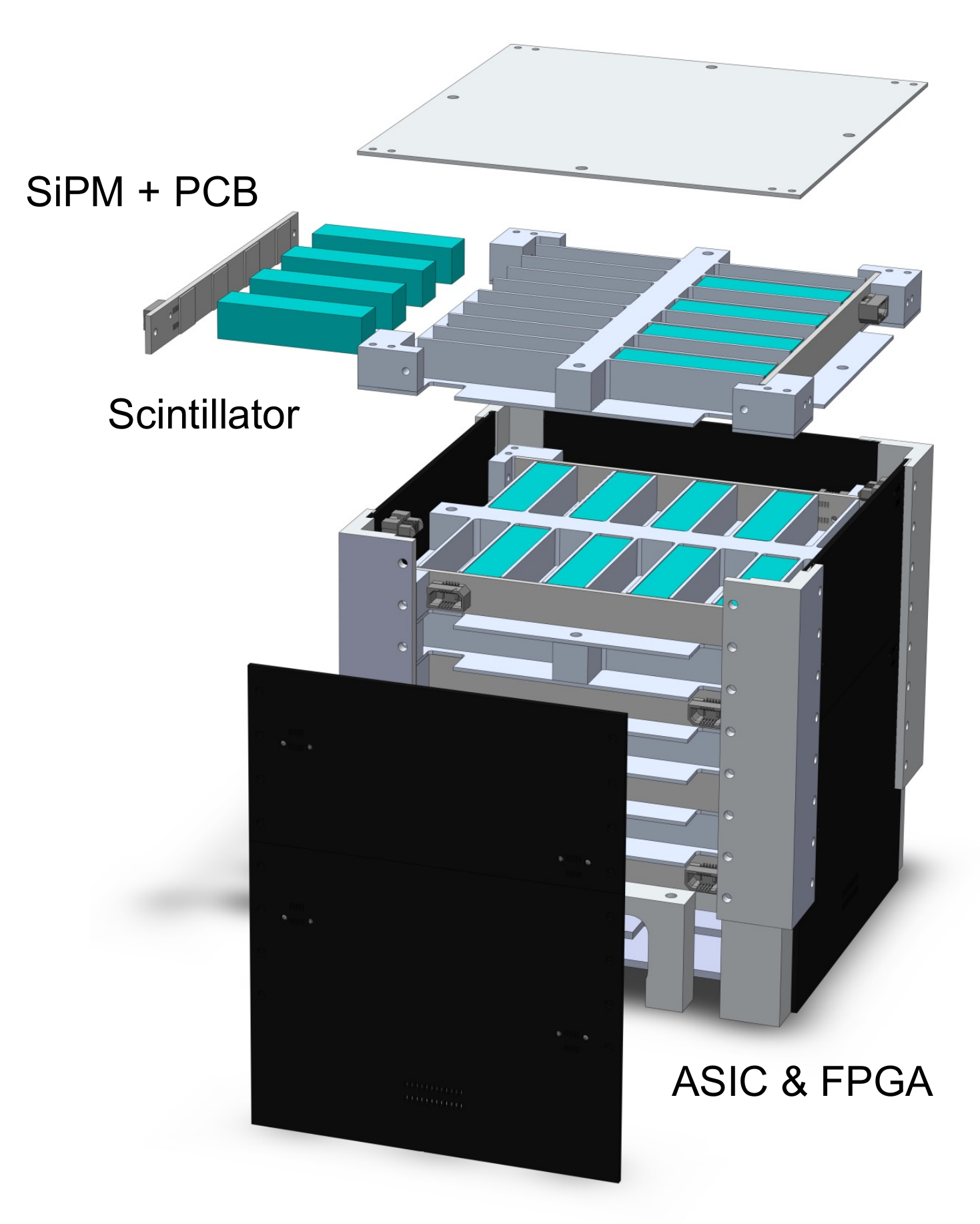}
      \end{center}
    \caption{The CAD drawing of the ComSAD detector. 
  \label{fig:structure}}
\end{figure}

In ComSAD, the 64 detection units (scintillator+SiPM pairs) are equally distributed into 8 layers, so that the 8 units in each layer are placed next to each other with an interval of ``one scintillator''. 
In order to optimize the number of the scintillators and the performance of the direction determination simultaneously, the orientations of scintillators in different layers are differed by $90^{\circ}$. 
The supporting structure of ComSAD is made from 6061 aluminium alloy (Al-6061) and includes 7 layers for housing scintillators, 1 top plate, 1 bottom layer for housing scintillator and PCBs, 4 side plates for SiPM, and 4 L-shape holders.  
Figure~\ref{fig:structure} shows the configuration of the scintillators in the ComSAD detector. 
The ComSAD assembly dimensions are 12.0 $\times$ 12.3 $\times$ 12.3 $cm^3$. 

\begin{figure}[!htbp]
  \begin{center}
      \includegraphics[width=0.42\textwidth]{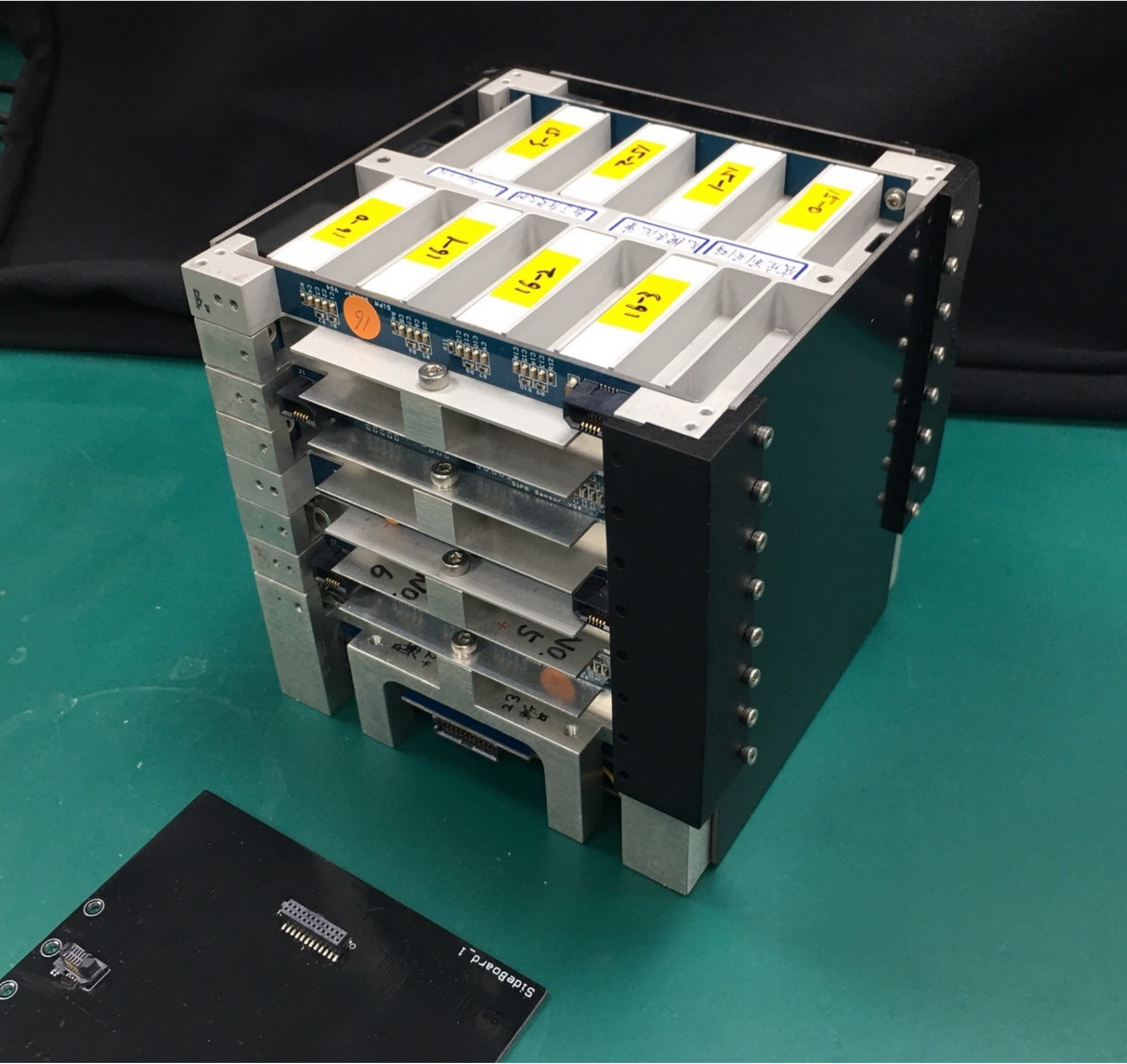}
      \end{center}
    \caption{An image of the assembled ComSAD detector. 
  \label{fig:comsad_lite}}
\end{figure}

To meet the requirements for the cost and weight of the ``forward-looking hybrid sounding rocket project'', the aforementioned key components for ComSAD are (1) the plastic scintillator (BC-408) from Saint-Gobain, which is sensitive to charge particles with the peak emission of light at 425 $nm$~\cite{bc408}. Each scintillator is painted by the reflector paint (BC-620)~\cite{bc620} to gain the better reflection from the sides. (2) The SiPM (C-series 6$\times$6 $mm^2$) from SensL, for which the peak wavelength of detection is 420 $nm$, the gain is 6 $\times 10^6$ and the photon detection efficiency is $\sim47$\%~\cite{sipm}. (3) The ASIC (SPIROC 2E) from OMEGA, which has 36 analog inputs~\cite{asic}. (4) The industrial-grade ASP1000-PQG208 FPGA~\cite{fpga}.
The circuit for ASIC is based on the SPIROC 2E test board designed by OMEGA. 
The power consumption of ComSAD is about 10 W and the sampling rate is 1 MHz.  
Figure~\ref{fig:comsad_lite} shows the assembled ComSAD detector. 

\section{Performance of ComSAD}
\begin{figure}[!htbp]
  \begin{center}
      \includegraphics[width=0.32\textwidth]{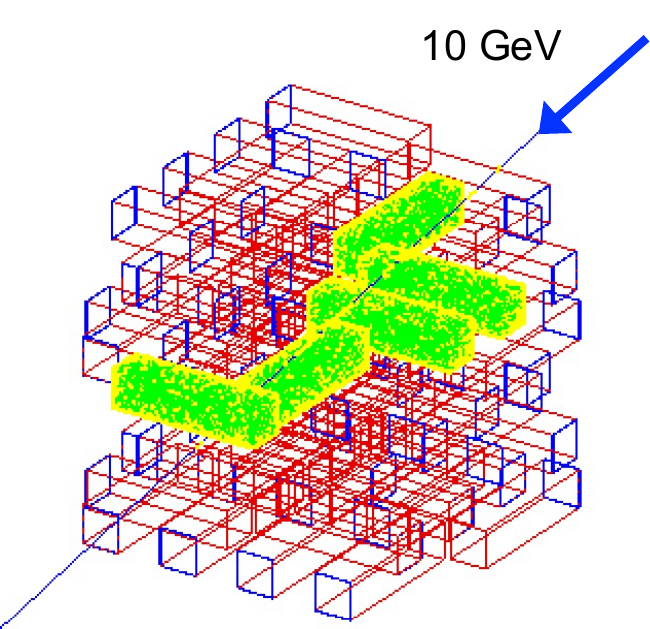}
      \end{center}
    \caption{The detector response of a proton with 10 GeV kinetic energy hitting on ComSAD.
  \label{fig:simulation}}
\end{figure}

\begin{figure}[!htbp]
  \begin{center}
      \includegraphics[width=0.4\textwidth]{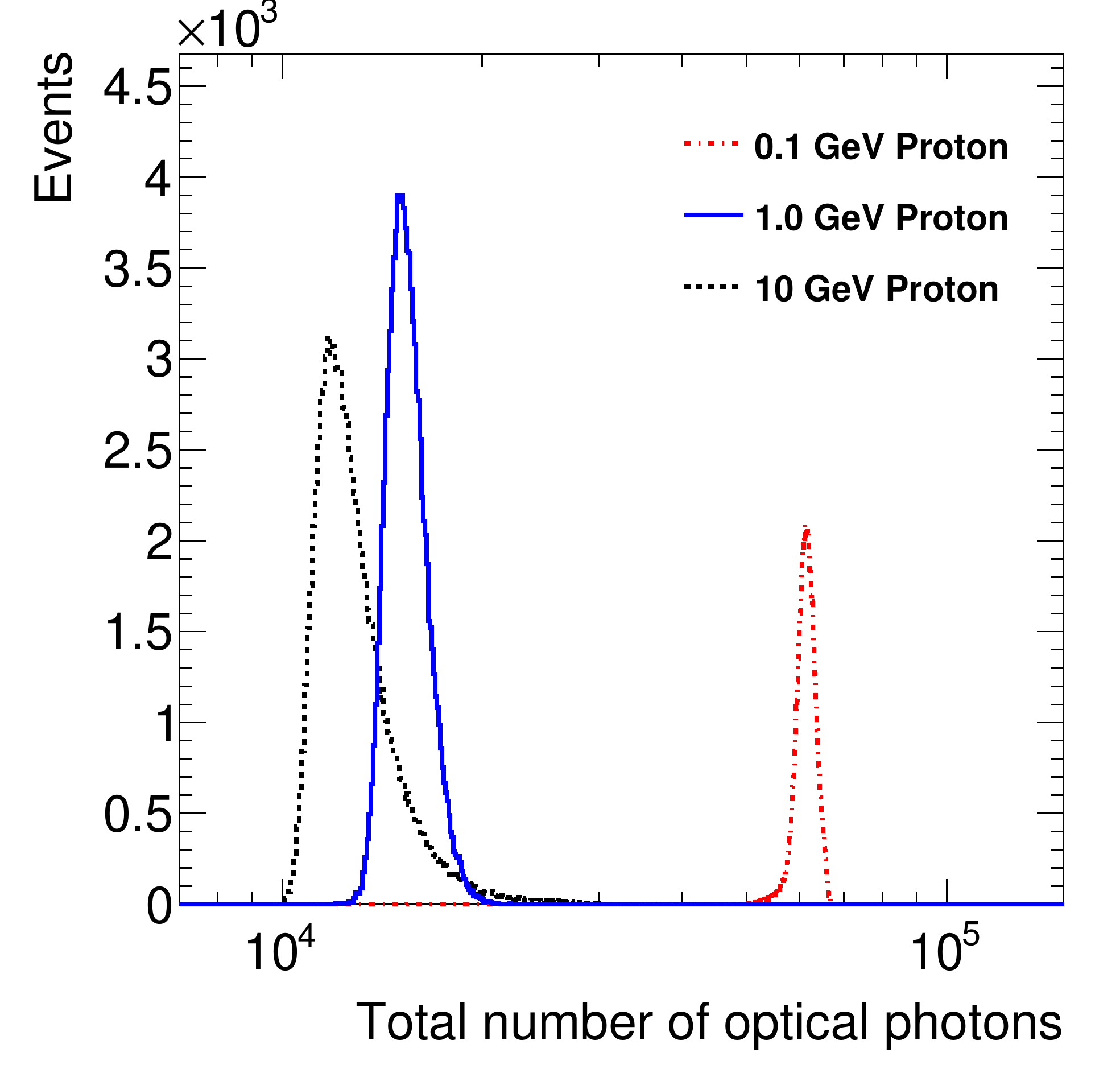}
      \end{center}
    \caption{The total number of optical photons hitting on the SiPM of 0.1 (red-dashed-dotted histogram), 1.0 (blue histogram), and 10 GeV (black-dotted histogram) protons. 
  \label{fig:n_photon}}
\end{figure}

The detector performance of ComSAD can be obtained from the GEANT4 simulations~\cite{geant4}, and using vendor-supplied specifications as input parameters for the scintillators, SiPM, and supporting structure. 
Figure~\ref{fig:simulation} shows an event display of ComSAD for an incident proton with 10 GeV kinetic energy, and the light-green color indicates the scintillators that would have detected a photon. 
The total numbers of optical photons hitting the SiPM with different energies of incident protons, 0.1, 1.0, and 10 GeV, are shown in Fig.~\ref{fig:n_photon} and it is clear that ComSAD has the capability of distinguishing the energy of incident particles. 

\begin{figure*}[!htbp]
  \begin{center}
     \subfigure[]{
      \label{fig:n_scintillators}
      \includegraphics[width=0.4\textwidth]{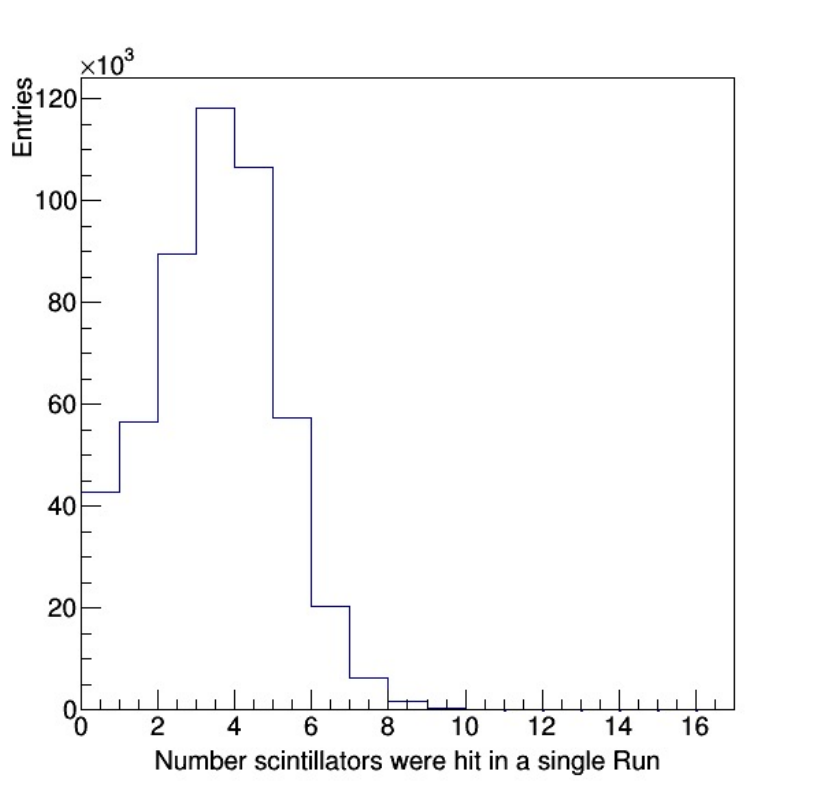}
    }
    \subfigure[]{
      \label{fig:n_channel}
      \includegraphics[width=0.4\textwidth]{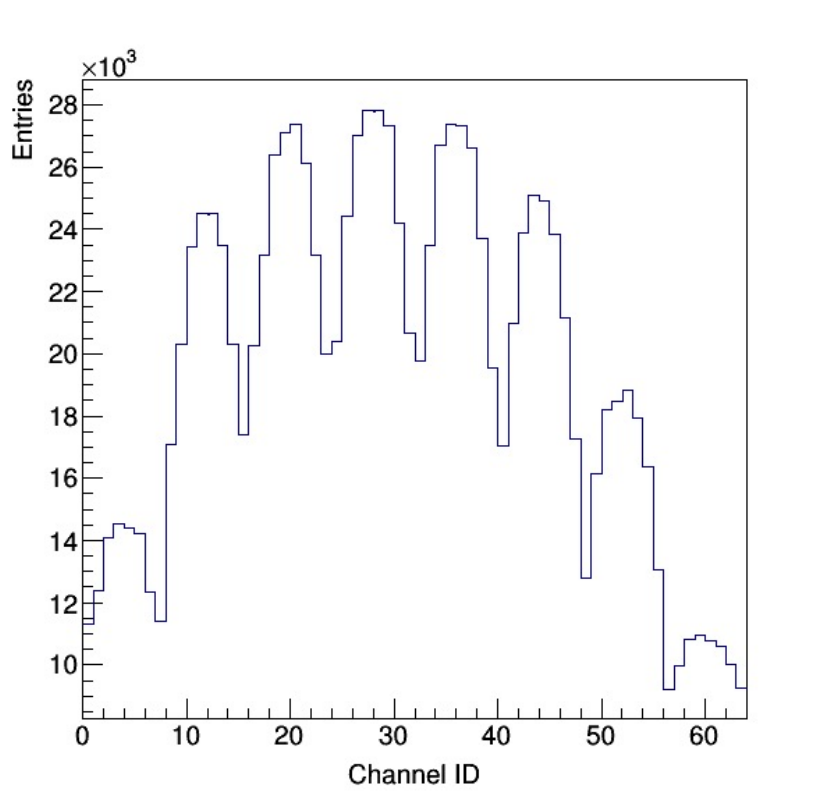}
    }  
  \end{center}
    \caption{ ~\subref{fig:n_scintillators} The distributions of the number of triggered scintillators and ~\subref{fig:n_channel} the distribution of triggered channels using 500 k randomly distributed events with 10 GeV protons. 
    \label{fig:sim2}}
\end{figure*}

Figures~\ref{fig:n_scintillators} and~\subref{fig:n_channel} show the number scintillators that saw an event and the distribution of readout channels resulting from a simulation that randomly generated 500k events with 10 GeV protons located 10 cm above the ComSAD instrument plane.
It shows that most of the events trigger more than 3 scintillators simultaneously. 
The structure of the triggered channels shown in Fig.~\ref{fig:n_channel} comes from the layout configuration of scintillators in ComSAD, as described in Sec.~\ref{sec:comsad}. 

\begin{figure}[!htbp]
  \begin{center}
      \includegraphics[width=0.4\textwidth]{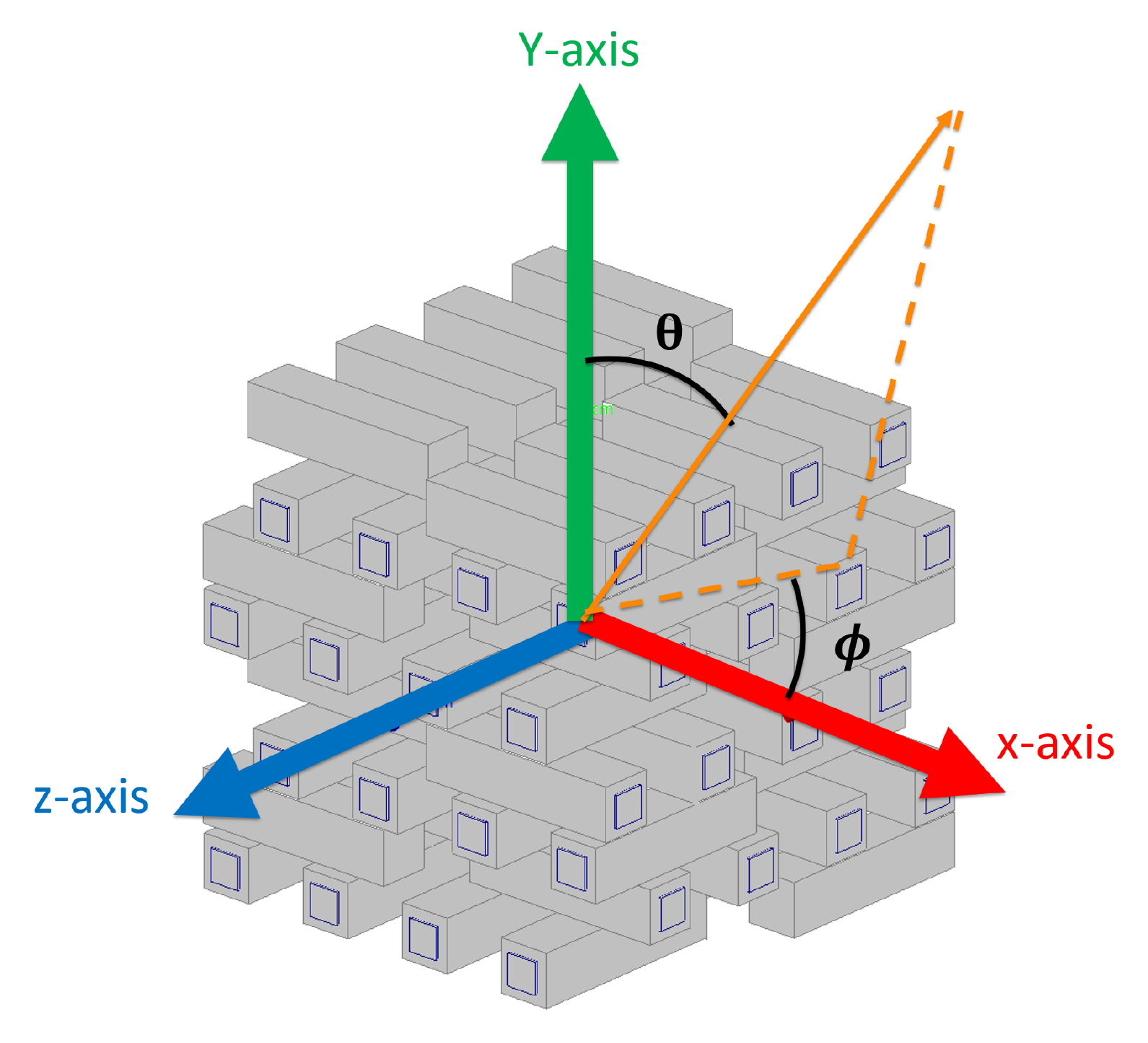}
      \end{center}
     \caption{The coordinate system for ComSAD that is used for tracking. The (0, 0, 0) is at the center of ComSAD, the $x-$axis ($z-$axis) is defined as along the long-sided of scintillators of odd (even) layers, and the y-axis is defined by using the right-handed rule.
  \label{fig:coord}}
\end{figure}

To determine the direction of the incident particles passing through ComSAD, at least three coincident fired scintillators in an event are needed. 
The origin (0, 0, 0) of the coordinate system is set at the center of ComSAD, the $x-$ and $z-$axis are defined as along the long-sided of scintillators of odd layers and even layers, respectively, and the y-axis is defined by using the right-handed rule, as shown in Fig.~\ref{fig:coord}. 

\begin{figure}[!htbp]
  \begin{center}
      \includegraphics[width=0.18\textwidth]{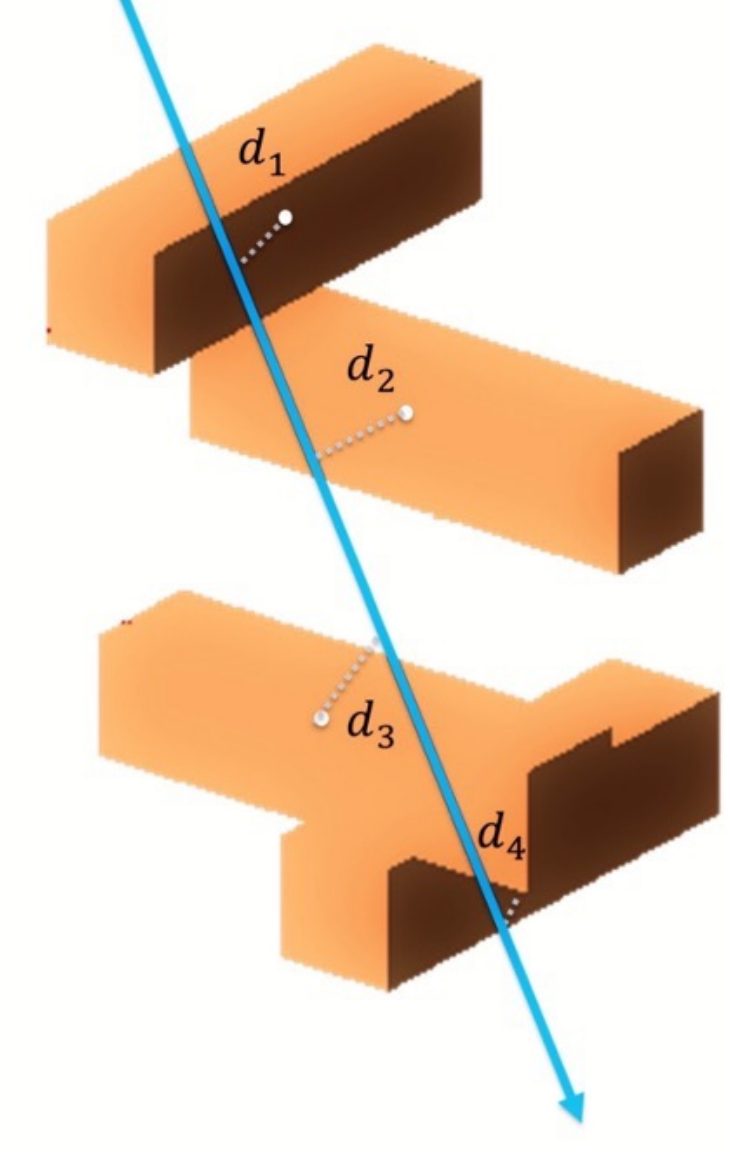}
      \end{center}
     \caption{A schematic showing the tracking in ComSAD. The length $h$ is the sum of the distance between the center of each fired scintillator and the track (blue arrow). 
  \label{fig:tracking}}
\end{figure}

A simple algorithm can be used to reconstruct the track of the incident particle. 
Firstly, the interaction position on each fired scintillator is set to be at the center of the scintillator volume. 
Secondly, an initial track, which is parameterized by a linear function, is guessed based on the top and bottom fired scintillators.
Thirdly, the total length ($h$) is defined as the sum of the distances between the center of each fired scintillator and the track (the linear function) as shown in Fig.~\ref{fig:tracking}, such that:
\begin{equation}
    \label{eq:distance}
    h =  \sum_{i} d_i = d_1 + d_2 + d_3 + d_4. 
\end{equation}
Finally, the optimal parameters of the linear function are determined by scanning the parameters with the different zenith angle $\theta$ and azimuthal angle $\phi$ within the range of $\pm 45^{\circ}$ with the minimum length $h$, which also corresponds to the maximum likelihood. 

\begin{figure*}[!htbp]
  \begin{center}
     \subfigure[]{
      \label{fig:2d_resol}
      \includegraphics[width=0.31\textwidth]{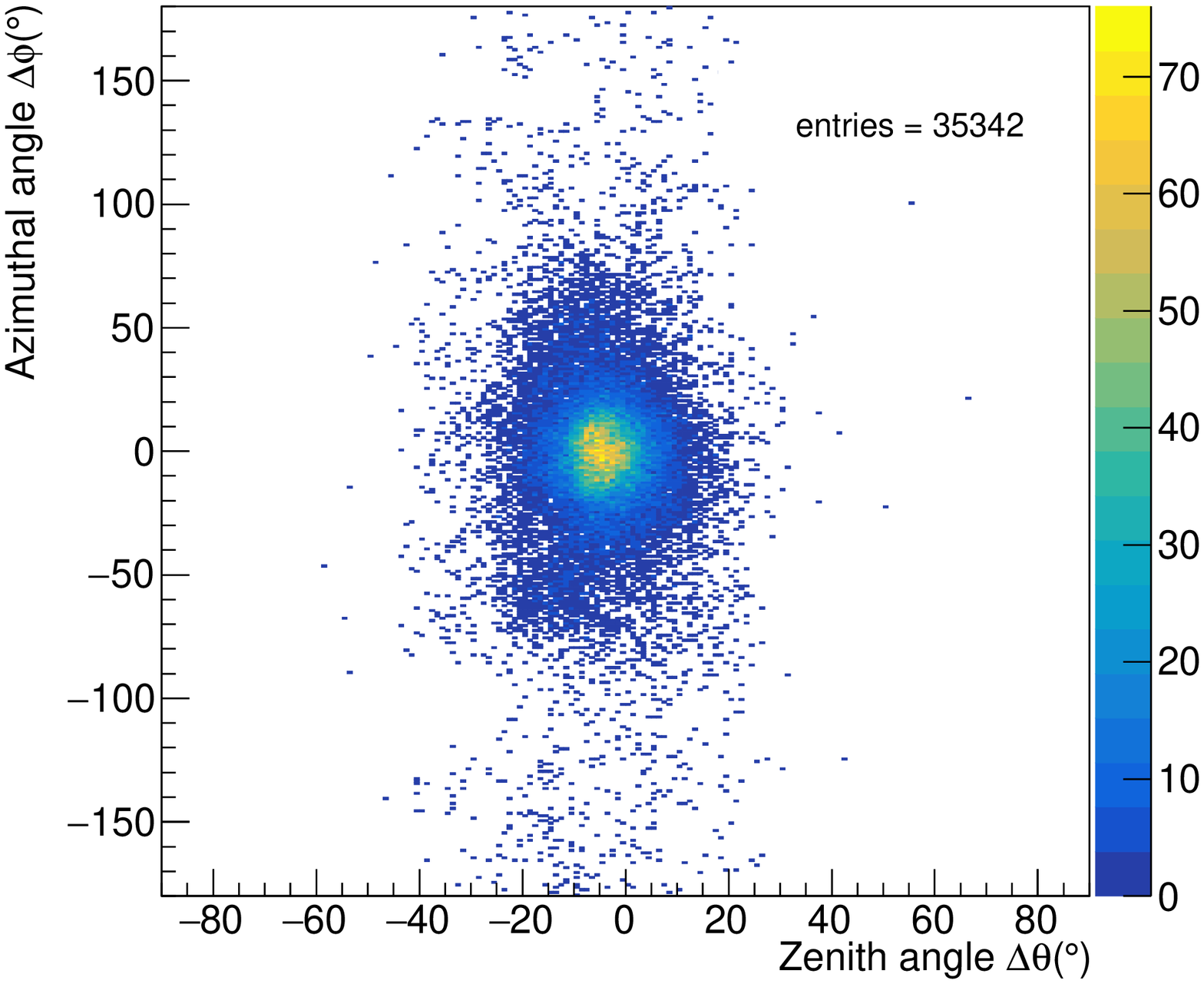}
    }
    \subfigure[]{
      \label{fig:dphi_resol}
      \includegraphics[width=0.31\textwidth]{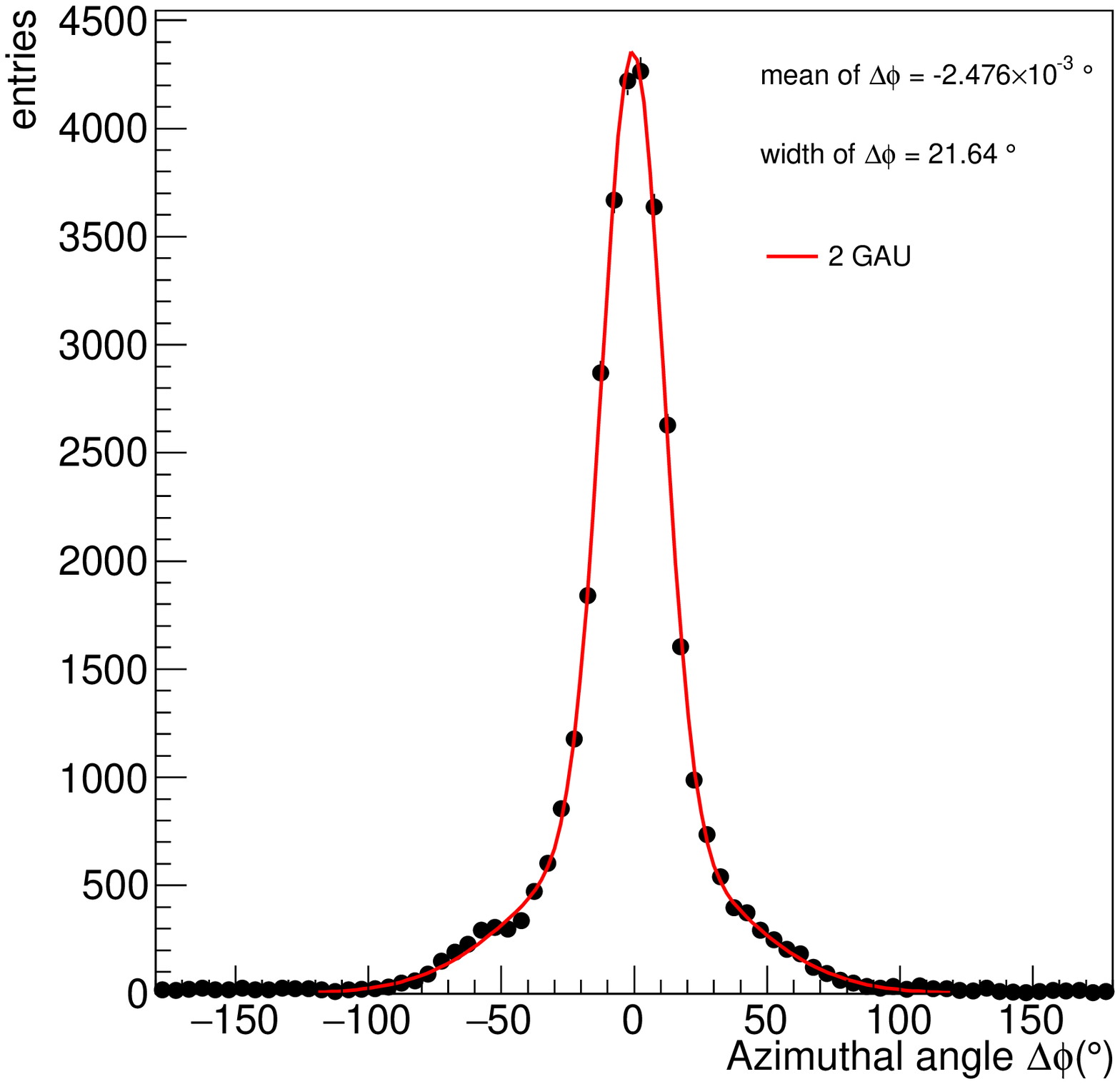}
    } 
    \subfigure[]{
      \label{fig:dtheta_resol}
      \includegraphics[width=0.31\textwidth]{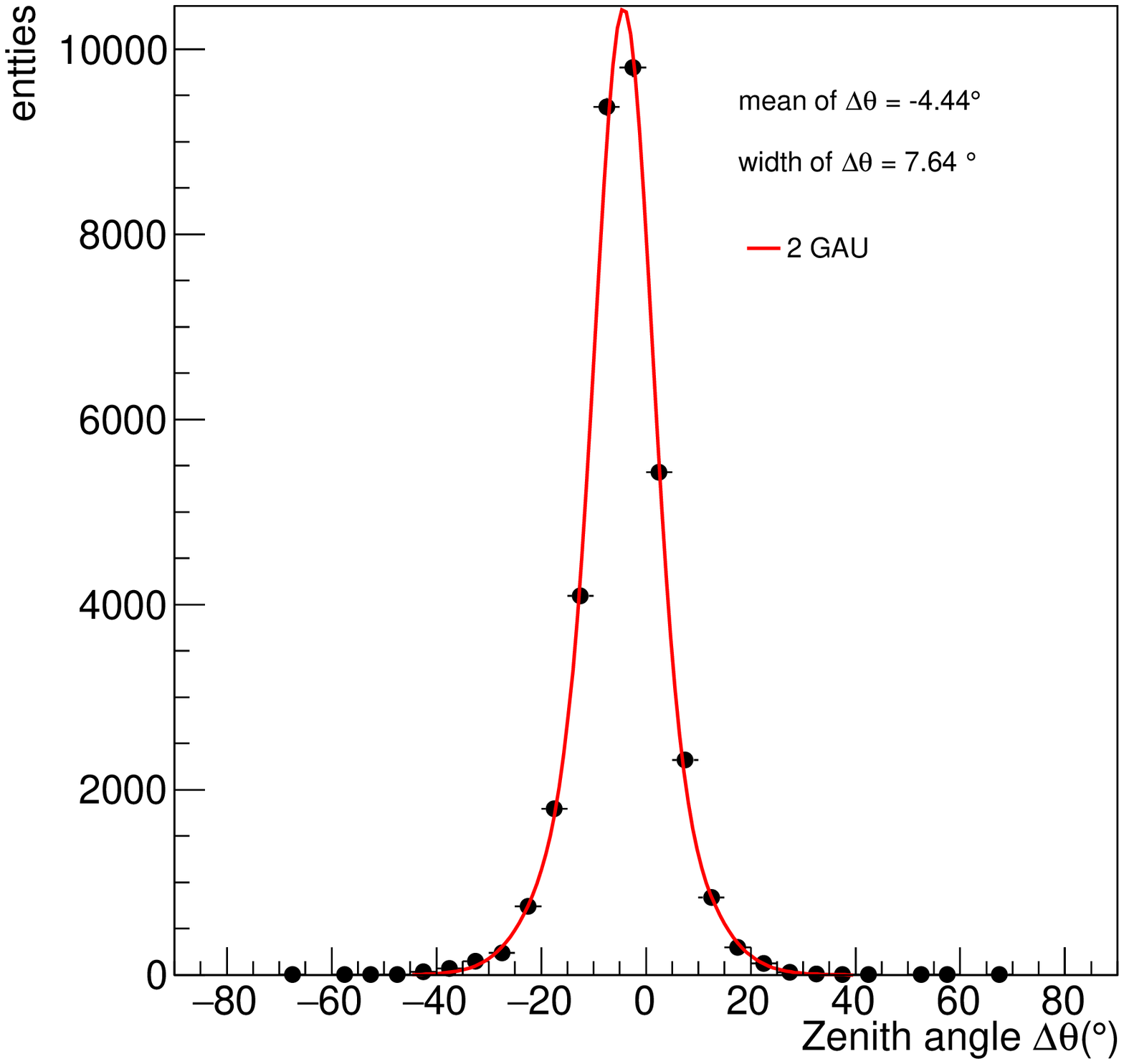}
    } 
  \end{center}
    \caption{ The angular resolutions \subref{fig:2d_resol} in 2 dimensions ($\phi$ v.s. $\theta$), \subref{fig:dphi_resol} in azimuthal angle $\phi$, and \subref{fig:dtheta_resol} in zenith angle $\theta$. 
    \label{fig:resol}}
\end{figure*}

Figures~\ref{fig:2d_resol} to~\ref{fig:dtheta_resol} show a comparison between the reconstructed and the input direction in 2 dimensions and in each angle $\phi$ and $\theta$, respectively.  
The results show that the bias in $\phi$ is $0.002^{\circ}$ and in $\theta$ is $-4.44^{\circ}$, while the resolutions in $\phi$ is $21.6^{\circ}$ and in $\theta$ is $7.64^{\circ}$.
The angular resolutions are also constrained by the layout configuration of scintillators in ComSAD. 
The better resolution in $\theta$ is due to more layers of scintillators, which means more information in the $y-$axis.

\begin{figure*}[!htbp]
  \begin{center}
     \subfigure[]{
      \label{fig:eff_fit}
      \includegraphics[width=0.4\textwidth]{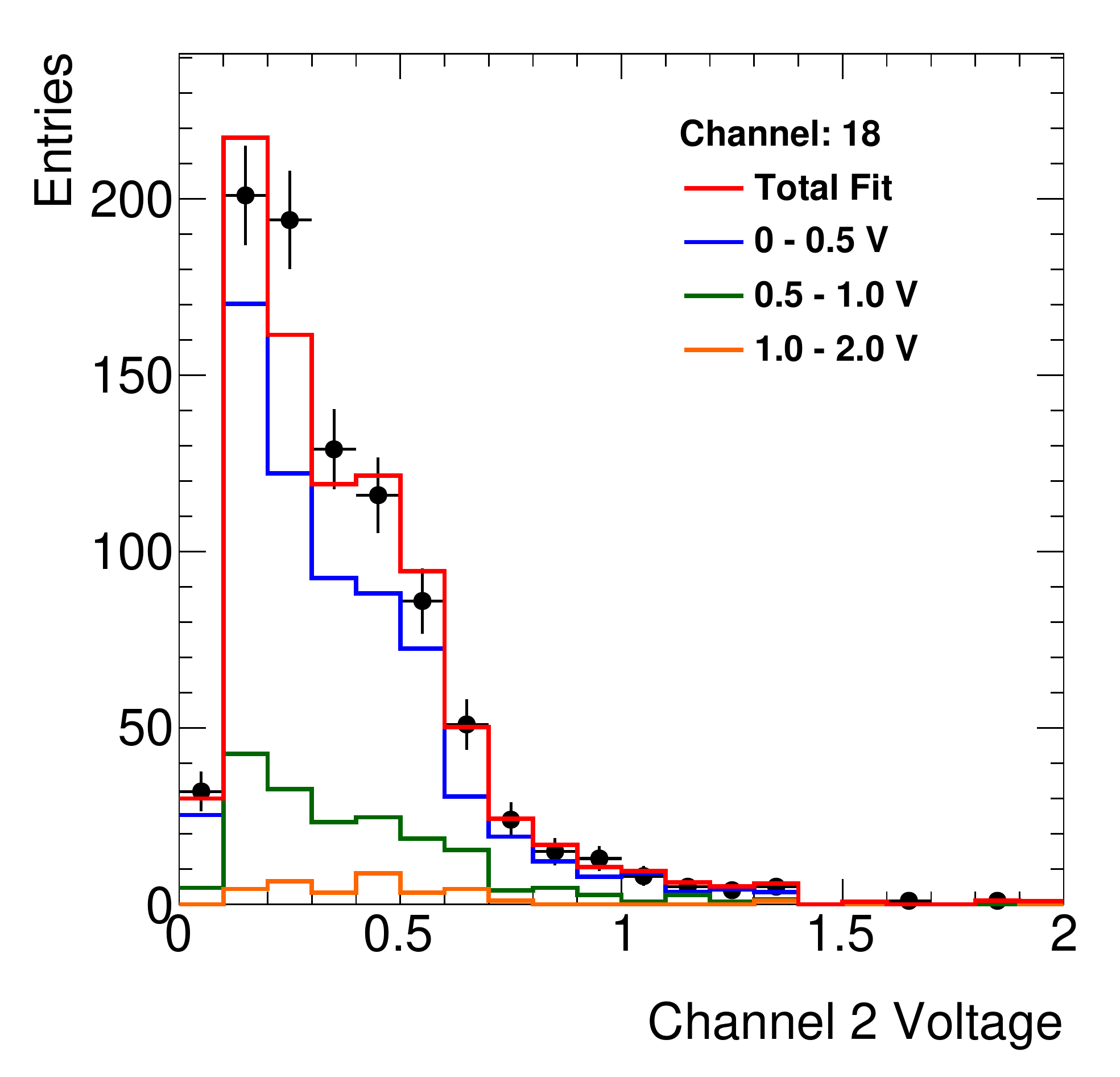}
    }
    \subfigure[]{
        \label{fig:eff}
      \includegraphics[width=0.4\textwidth]{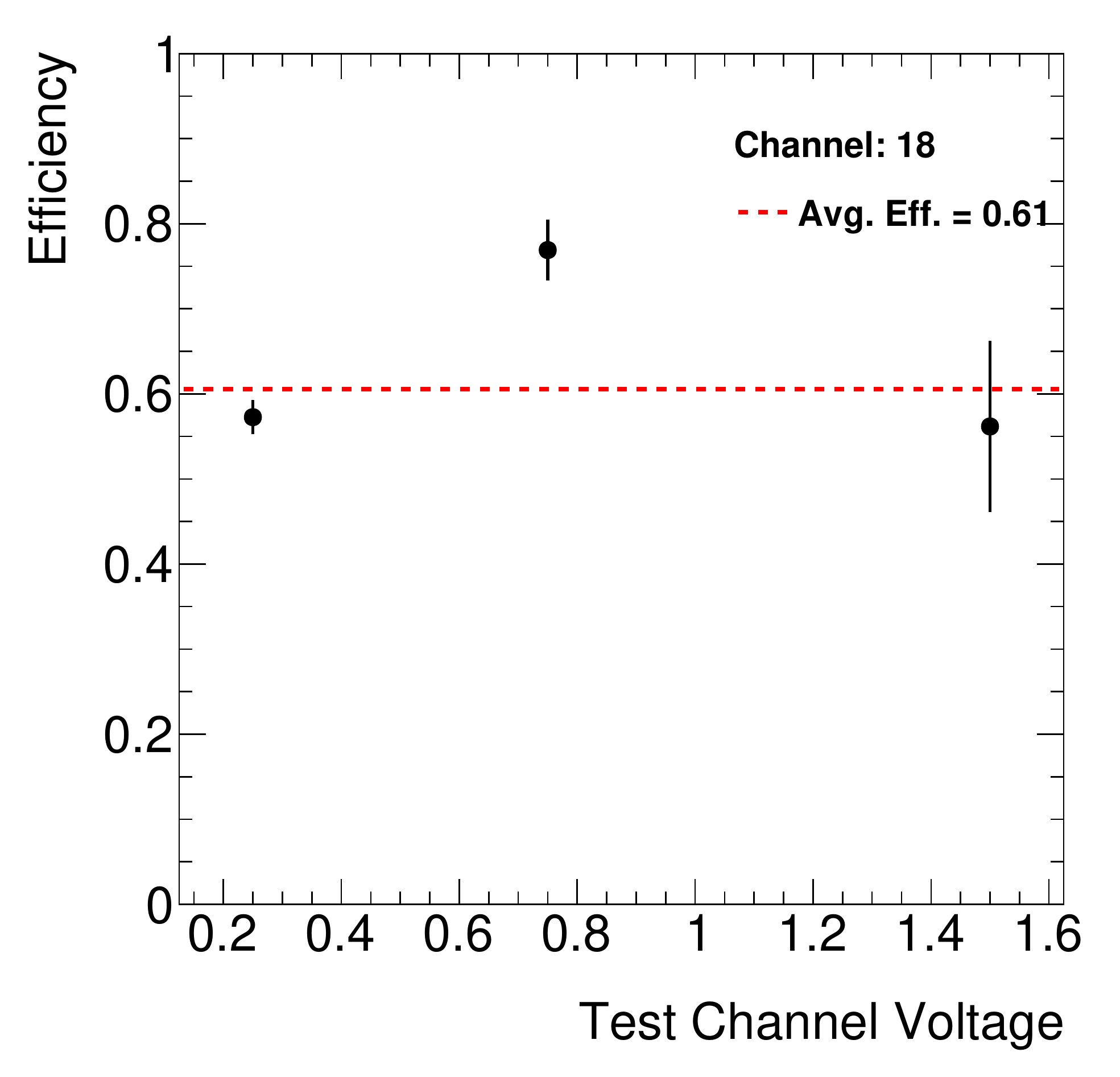}
    } 
  \end{center}
    \caption{  \subref{fig:eff_fit} The example for the fit result of detection efficiency. The data points (black points) are fitted by the templates obtained from the events that triggered channels 2 and T in certain V bins of T (the blue histogram is 0 - 0.5 V, the green histogram is 0.5 -1.0 V, the orange histogram is 1.0 - 2.0 V, and the red histogram is the total fit). \subref{fig:eff} The detection efficiency as a function of V of the test sample. 
    \label{fig:eff_result}}
\end{figure*}

It is also very important to understand the detection efficiency for each detection unit needed to provide a correct interpretation of our results.
Large statistics of cosmic-ray events are used to determine the detection efficiency. 
In the efficiency determination setup, there are two detection units serving as triggers, which are denoted as ``1'' and ``2'', and one test unit, which is denoted as``T'', that are sandwiched between two trigger units. 
The efficiency as a function of energy (voltage from the reading of SiPM) is defined as
\begin{equation}
    \label{eq:eff}
    \epsilon^{V} = \frac{N^V_{1,T,2}}{N^V_{1,2}},
\end{equation}
where $V$ denotes a specific voltage (energy) range for the T channel, $N^V_{1,T,2}$ is the number of events in which all units (1, 2, T) trigger simultaneously, and $N^V_{1,2}$ is the number of events in which channels 1 and 2 are triggered simultaneously, no matter if T is fired or not. 
$N^V_{1,T,2}$ can be obtained in a straightforward way, but $N^V_{1,2}$ can't due to the lack of information from the T channel. 
To extract $N^V_{1,2}$, the events in which the 1 (or 2) and T channels are triggered simultaneously are used. 
Firstly, the normalized V distribution of channel 1 (or 2) in different V bins of channel T are obtained as the templates. 
Secondly, these templates are used to fit the events in which channels 1 and 2 are triggered to obtain $N^V_{1,2}$ for different V bins of the T channel. 
Then, the efficiency is calculated using Eq.~\ref{eq:eff}. 
Figures~\ref{fig:eff_fit} and~\ref{fig:eff} show an example of the fit result and the detection efficiency as a function of voltage on the test sample, respectively. 
The result shows small energy dependence on the detection efficiency and the average efficiency is about 61\%.  

Table~\ref{tab:comsadspec} summarizes the specifications for ComSAD. 
The energy calibration will be described in a separate article as the primary goal in this stage of the ComSAD effort is to measure the flux and directions of incident cosmic rays.  



\begin{table}[!htbp]
\centering
\caption{ComSAD specifications.}
\begin{tabular}{|l|c|}\hline
    Dimension &12.0 $\times$ 12.3 $\times$ 12.3 $cm^{3}$ \\\hline
    Mass &1.72 kg \\\hline
    Power consumption & 10 W \\\hline
    Number of channels & 64 \\\hline
    $\theta$ bias & 0.002$^{\circ}$ \\\hline
    $\theta$ resolution & 7.64$^{\circ}$ \\\hline
    $\phi$ bias & -4.44$^{\circ}$ \\\hline
    $\phi$ resolution& 21.64$^{\circ}$ \\\hline
    Energy resolution & $\sim$10 GeV \\\hline
    Time resolution & $\sim$1 ms \\\hline
    Detection efficiency & $\sim$60\% \\\hline
\end{tabular}
\label{tab:comsadspec}
\end{table}

\section{Portable ComSAD}
The small footprint and low-power consumption of the cosmic-ray detector ComSAD make it ideal for adaption on to an aircraft platform.
Modifications needed to do this include using a Raspberry PI as the onboard computer for controlling the payload, adding a battery and DC-DC converter in the PSU, adding the GNSS receiver to record the position information, and adding an interface for user instrument control.  

\begin{figure*}[!htbp]
  \begin{center}
      \includegraphics[width=0.9\textwidth]{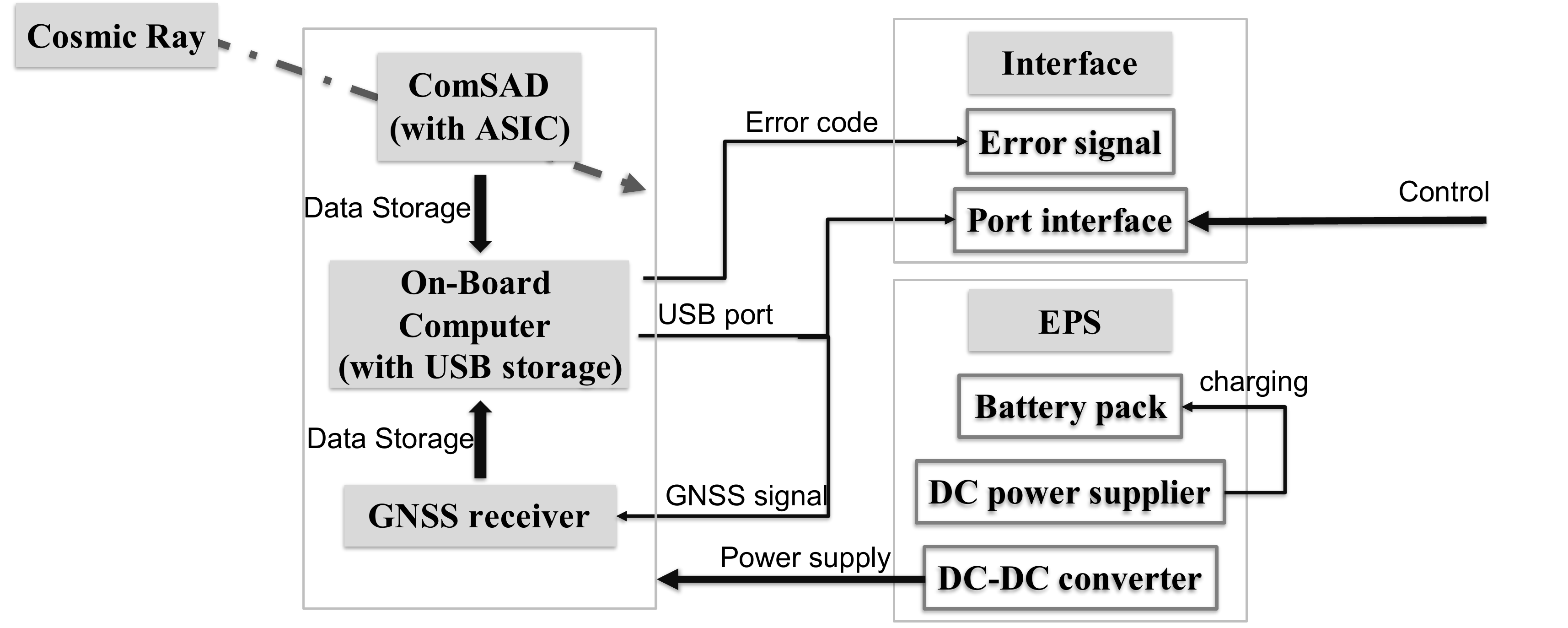}
      \end{center}
    \caption{pComSAD systems architecture block diagram.  
  \label{fig:port_comsad_sys}}
\end{figure*}

\begin{figure}[!htbp]
  \begin{center}
      \includegraphics[width=0.4\textwidth]{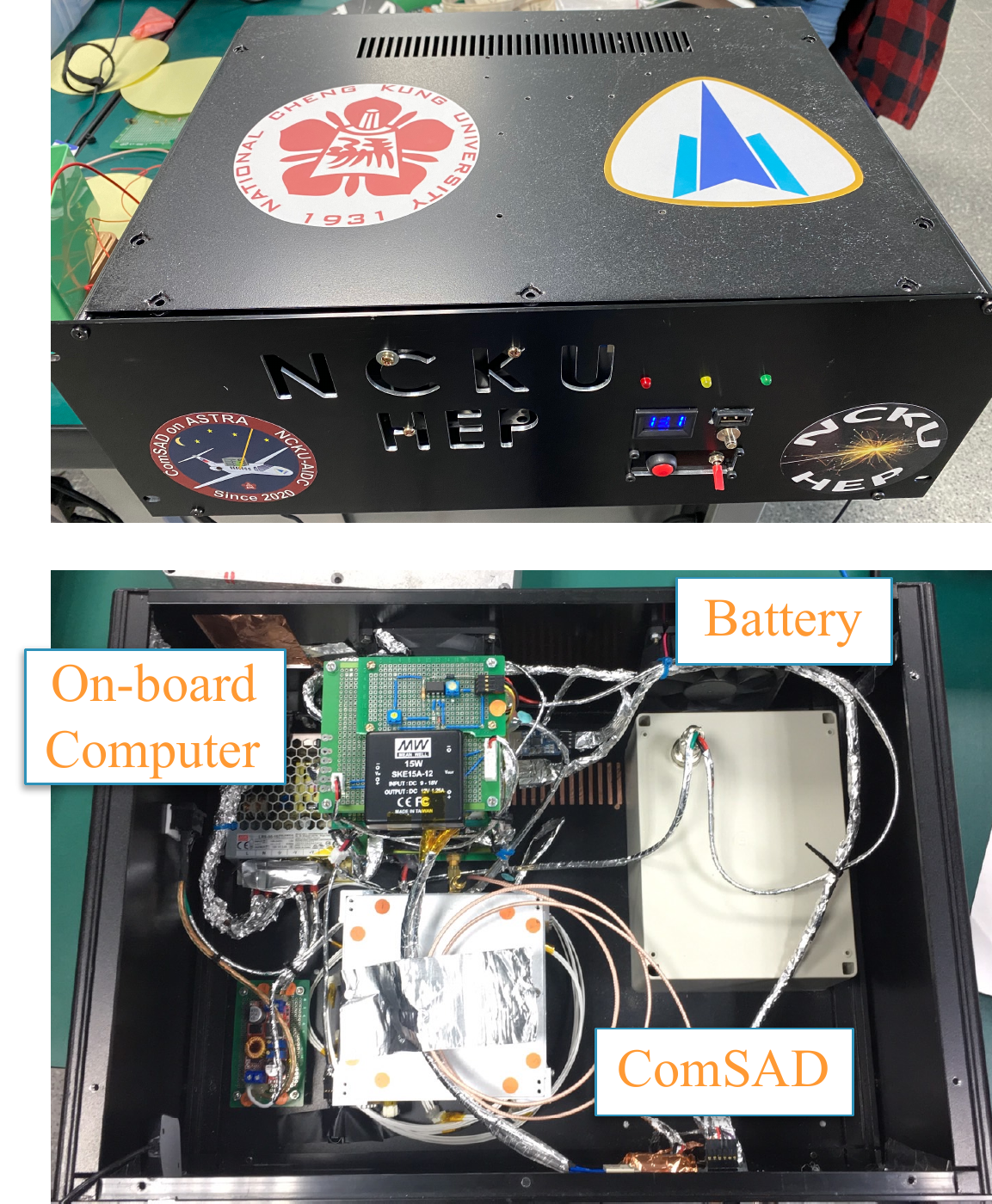}
      \end{center}
    \caption{An image of the pComSAD instrument.
  \label{fig:port_comsad}}
\end{figure}

This modified version of ComSAD is called ``portable ComSAD'' or ``pComSAD''.  
The system architecture and instrument are shown in Figs.~\ref{fig:port_comsad_sys} and~\ref{fig:port_comsad}, respectively.  
The details of pComSAD are found in Ref.~\cite{pk_thesis, nina_thesis}. 

\section{Conclusions}
While many ground-, underground-, and Earth orbiting-based experiments are dedicated to measuring cosmic rays with high precision, there is still quite a bit of room for small and low-cost experiments to make an impact. 
This is, in part, due to recent developments of CubeSats and an increased number of opportunities for launching sounding rockets.

Compact Scintillator Array Detector, ComSAD, is a scintillator-based cosmic-ray detector that is funded under the ``forward-looking hybrid sounding rocket project'' from NSPO in Taiwan. 
It consists of 64 channels in total and can provide flux, direction, and energy measurements of cosmic rays. 
The angular resolutions is around $10-20^{\circ}$ in each direction and the detecting efficiency for each unit is about 60\% without obvious energy dependency. 

ComSAD can also be modified for the future CubeSat and aircraft missions to collect more cosmic-ray data at different altitudes and locations. 
However, there is still some room for further improvements on energy and angular resolutions by modifying the configurations of the scintillators. 
This first version of ComSAD provides us the basic foundation for further work.  
Finally, ComSAD is the first ``made-in-Taiwan'' cosmic-ray detector used in sounding rocket and CubeSat missions and it will open a new door for cosmic ray physics. 

\section*{Acknowledgments}
We thank National Cheng Kung University and National Space Organization, Taiwan, R.O.C. for their support. 
This work is also supported by the Ministry of Science and Technology, Taiwan, R.O.C. 
We also thank Aerospace Industrial Development Corporation (AIDC) and Dr. Stéphane Callier from OMEGA for the useful suggestions and technical supports.  



\begin{thebibliography}{99}

\bibitem{cr_hist_1}
    {V. F. Hess}, Uber beobachtungen der durchdringenden strahlung bei sieben freiballonfahrten, Phys. Zeits. {\bf 13} (1912) 1084.

\bibitem{cr_hist_2}
    {T. Wulfetal}, Observations on the radiation of high penetration power on the Eiffel tower, Physikalische Zeitschrift {\bf 11} (1910) 811.

\bibitem{icecube}
    {M. G. Aartsen {\it et al.} (IceCube Collaboration)}, Evidence for high-energy extraterrestrial neutrinos at the IceCube detector, Science {\bf 342} (2013) 1242856.

\bibitem{hegra} 
    {R. Mirzoyan {\it et al.} (The HEGRA Collaboration)}, The first telescope of the HEGRA air Cherenkov imaging telescope array, Nucl. Instr. and Meth. A {\bf 351} (1994) 513.  

\bibitem{cream} 
    {E.S. Seo {\it et al.}}, Cosmic-ray energetics and mass (CREAM) balloon project, Advances in Space Research {\bf 33} (2004) 1777.

\bibitem{gaps} 
    {C J Hailey {\it et al.} (GAPS Collaboration)}, Accelerator testing of the general antiparticle spectrometer; a novel approach to indirect dark matter detection, JCAP {\bf 01} (2006) 007.

\bibitem{ams} 
    {M. Aguilar {\it et al.} (AMS Collaboration)}, First Result from the Alpha Magnetic Spectrometer on the International Space Station: Precision Measurement of the Positron Fraction in Primary Cosmic Rays of 0.5 – 350 Gev, Phys. Rev. Lett. {\bf 110} (2013) 141102. 

\bibitem{pamela} 
    {P. Picozza {\it et al.} (PAMELA Collaboration)}, A payload for antimatter matter exploration and light-nuclei astrophysics, Astropart. Phys. {\bf 27} (2007) 296.

\bibitem{rocket} 
    NASA Sounding Rocket Program Handbook, (2015).

\bibitem{gw_exp}
    {B. Abbott  {\it et  al.} (LIGO  Scientific Collaboration and VirgoCollaboration)}, Observation of Gravitational Waves from a Binary Black Hole Merger, Phys. Rev.  Lett.  {\bf 116} (2016) 061102. 
  
\bibitem{cubesat} 
   {Cal Poly SLO} The CubeSat Program, CubeSat Design Specification Rev.13, (2014).
  
\bibitem{nspo} 
    {\url {https://www.nspo.narl.org.tw}}
  
\bibitem{bc620} 
    {\url {http://static6.arrow.com/aropdfconversion/5f21e5069a452c0b4d6a95c9954f098ba0ad6261/113423053200175sgc-bc620-data-sheet.pdf}}
  
\bibitem{bc408} 
    {\url {https://www.crystals.saint-gobain.com/products/bc-408-bc-412-bc-416}}
  
\bibitem{sipm} 
    {\url {https://sensl.com}}
  
\bibitem{asic} 
    {\url {https://portail.polytechnique.edu/omega/en/products/products-presentation/spiroc}}
  
\bibitem{fpga}
    {\url {https://www.mouser.tw/datasheet/2/268/microsemi_proasic3_flash_family_fpgas_datasheet_ds-1592395.pdf}}
  
\bibitem{geant4} 
    {S. Agostinelli {\it et  al.}}, GEANT4 - a simulation toolkit, Nucl. Instr. and Meth. A {\bf 506} (2003) 250.
  
\bibitem{pk_thesis} 
    {P. K. Wang}, The R\&D of Compact Scintillator Array Detector for Cosmic Ray measurement in sounding rocket or CubeSat missions, {\url {http://ir.lib.ncku.edu.tw/handle/987654321/182841}} (2018).

\bibitem{nina_thesis} 
    {C.-Y. Chen}, Measuring Cosmic Rays at Different Altitude Ranges with Compact Scintillator Array Detector, (2021).

\end{thebibliography}
\end{document}